\documentclass[journal]{IEEEtran}
\IEEEoverridecommandlockouts
\usepackage{amsmath,amssymb}
\usepackage{cancel}
\usepackage{epsfig}
\usepackage{setspace}
\usepackage[linesnumbered,ruled]{algorithm2e}
\usepackage{algpseudocode}
\usepackage{amsmath}
\usepackage{color}
\usepackage{graphics}
\usepackage{float}
\usepackage{array}
\usepackage{booktabs}
\usepackage{bm}
\usepackage{multirow}
\usepackage{url}
\usepackage{cite}
\usepackage{CJK}
\usepackage{subfigure}
\usepackage[export]{adjustbox}
\usepackage{graphicx}

\makeatletter

\newcommand{\Rmnum}[1]{\expandafter\@slowromancap\romannumeral #1@}

\makeatother

\begin{document}

\title{Domain Adaptation and Autoencoder Based Unsupervised Speech Enhancement}

\author{Yi~Li,~\IEEEmembership{ Student~Member,~IEEE,}
        ~Yang~Sun,~\IEEEmembership{ Member,~IEEE,}\\
        Kirill~Horoshenkov,~\IEEEmembership{ Senior~Member,~IEEE,}
        and~Syed Mohsen~Naqvi,~\IEEEmembership{Senior~Member,~IEEE}
\thanks{Y. Li and S. M. Naqvi are with the Intelligent Sensing and Communications Group, School of Engineering, Newcastle University, Newcastle upon Tyne NE1 7RU, U.K. (e-mails: {y.li140, mohsen.naqvi}@newcastle.ac.uk)

Y. Sun is working with the Big Data Institute, University of Oxford, Oxford OX3 7LF, U.K. (e-mail: Yang.sun@bdi.ox.ac.uk)

K. Horoshenkov holds the position of a Personal Chair at the University of Sheffield. (e-mail: k.horoshenkov@sheffield.ac.uk)
}



}




\maketitle

\begin{abstract}
As a category of transfer learning, domain adaptation plays an important role in generalizing the model trained in one task and applying it to other similar tasks or settings. In speech enhancement, a well-trained acoustic model can be exploited to obtain the speech signal in the context of other languages, speakers, and environments. Recent domain adaptation research was developed more effectively with various neural networks and high-level abstract features. However, the related studies are more likely to transfer the well-trained model from a rich and more diverse domain to a limited and similar domain. Therefore, in this study, the domain adaptation method is proposed in unsupervised speech enhancement for the opposite circumstance that transferring to a larger and richer domain. On the one hand, the importance-weighting (IW) approach is exploited with a variance constrained autoencoder to reduce the shift of shared weights between the source and target domains. On the other hand, in order to train the classifier with the worst-case weights and minimize the risk, the minimax method is proposed. Both the proposed IW and minimax methods are evaluated from the VOICE BANK and IEEE datasets to the TIMIT dataset. The experiment results show that the proposed methods outperform the state-of-the-art approaches.
\end{abstract}

\textbf{\textit{Impact Statement-}Speech enhancement plays an essential role in real-world applications such as teleconferencing. However, unsupervised learning is challenging to realize but vital in unknown speech environments. This paper facilitates the domain adaptation research in unsupervised speech enhancement. In particular, we propose the importance-weighting and minimax methods to further improve speech enhancement performance. This work will help developers to save computational cost when applying to different testing groups. The proposed methods are also beneficial for researchers in other transfer learning tasks such as transferring a model trained for one language to another.}\\
\\

\IEEEpeerreviewmaketitle

\section{INTRODUCTION}

\IEEEPARstart{I}{n} recent years, machine learning research has been developed and exploited in speech enhancement. In order to solve the tasks in real-world applications such as hearing aids, machine translation, and robotics, various techniques, including deep learning, reinforcement learning (RL), and transfer learning, have been extensively utilized for the past decade \cite{Student}. As the main concept of deep learning, it refers to the hidden layers and neural units of various network models that have been applied in supervised and unsupervised problems. It has significantly improved speech enhancement performance because of the regression model \cite{EU,CSA1,Two}. The main target for RL algorithms is to decide a direction or action in different environments for maximizing the sum of a cumulative reward \cite{RL}. Due to the unique principle, the RL has been exploited in computer game design and dialogue management \cite{RL1}.

As one of the categories, transfer learning has been extensively utilized in speech enhancement to reduce computational complexity  \cite{TL}\cite{TL1}. In recent studies, because all human languages share some common semantic structures, transfer learning has been proposed to adopt a well-trained neural network model for crossing various settings, including languages, speakers, genders, and environments \cite{adapt}. In the multi-domain problems, the main concept of transfer learning is to build the domains crossing correspondence by the shared classes, and the model trained in one domain is transferred and reused in different domains. Xu et al. exploited deep neural networks (DNNs) obtained with high-resource materials for one language to cross over to another target language using a small amount of adaptation data \cite{cross}. Furthermore, some unseen speaker and noise problems were studied and the performance was improved by speech enhancement generative adversarial networks (SEGANs) \cite{crossnoise}.

\begin{table*}[htbp!]
\caption{SPEECH ENHANCEMENT PERFORMANCE COMPARISON IN TERMS OF {\bfseries \upshape{STOI (IN $\%$)}} WITH {\bfseries VCAE} BUT DIFFERENT SNR LEVELS AND NOISES. THE {\bfseries VOICE BANK} DATASET IS USED IN THE TRAINING STAGE AND DIFFERENT DATASETS ARE FOR THE TESTING STAGE. {\bfseries BOLD} INDICATES THE BEST RESULTS.}
\centering
\begin{tabular}{|c|c|c|c|c|c|c|c|c|c|c|}
\hline
Noise  & \multicolumn{3}{|c|}{psquare} & \multicolumn{3}{|c|}{living} & \multicolumn{3}{|c|}{station}&Average\\
\hline
SNR level (dB) & -5  & 0 & 5 & -5  & 0  & 5 & -5  & 0  & 5 &  \\     
 \hline
 Unprocessed & 58.5 & 61.3& 66.6& 57.3& 59.2& 63.8& 56.8&59.7& 60.0& 60.1 \\
 \hline
VOICE BANK \cite{VB}  &{\bfseries 76.1} &{\bfseries 78.5}  &{\bfseries 82.8}&{\bfseries 71.5}&{\bfseries 73.6}&{\bfseries 79.9}&{\bfseries 71.8}&{\bfseries 73.0}&{\bfseries 76.8} &{\bfseries 76.0} \\
 \hline
IEEE \cite{IEEE}  &73.8 & 77.1 & 82.5 &70.6 & 72.9 & 77.5&69.3 & 71.6 & 75.4&74.5 \\
 \hline
TIMIT \cite{TIMIT}   &73.4 & 76.9 & 82.1 &70.5 & 72.6 & 77.0&69.4 & 71.3 & 74.9&74.2\\
 \hline
WSJ0 \cite{WSJ0}   &72.1 & 75.3 & 80.9 &69.4 & 71.2 & 75.8&68.8 & 70.9 & 75.0&73.2\\
 \hline 
\end{tabular}
\end{table*}
Domain adaptation plays an important role as a research aspect of transfer learning in modern applications such as automatic speech recognition (ASR), machine translation, and text classification \cite{VB,IEEE,TIMIT,WSJ0}. For example, Park et al. realized the robustness in ASR systems with generative adversarial networks (GANs) and disentangled representation learning \cite{adapt1}. In recent years, domain adaptation has become a highly studied task in speech enhancement due to the importance that the well-trained model is suitable for various scenarios. In \cite{adapt}, transfer component analysis (TCA) was proposed to solve the semi-supervised domain adaptation problems with maximum mean discrepancy (MMD). As a robust approach to the domain adaptation problems, domain adversarial training (DAT) extracts the domain invariant features and trains the discriminator to determine the input source based on the extracted features \cite{noisead}. Therefore, the information of the domains is not fully exploited in the downstream task and the above techniques have limitations. Besides, joint distribution adaptation (JDA) jointly utilizes both the marginal distribution and conditional distribution, and integrate JDA with Principal Component Analysis (PCA) to build new feature representation \cite{JDA}. The difference of both distributions between source and target domains is minimized by reducing MMD. However, the strict conditions of use and complex training process limit the JDA methods. Moreover, JDA methods require more labeled drift samples to participate in the model construction. Transfer learning becomes more challenging as domains may change by the joint distributions of input features and output labels, which is a common scenario in practical applications \cite{JDA2}.


In order to improve the performance, neural networks such as recurrent neural networks (RNNs) and GANs have been introduced in domain adaptation problems \cite{CVPR}\cite{IWA}. For example, Zhang et al. exploited importance weighted adversarial networks within the partial domain adaptation approach and reduced the shift between the target data and the source data \cite{IWGAN}. The cross-domain method has a substantially better performance in a specific scenario, where it is required to transfer from a larger and more diverse source domain to a smaller and more similar target domain with less number of classes. However, in real-world scenarios, a well-trained model is generally required for a more challenging case such as transferring the model from less and dependent source speakers to more and independent speakers \cite{noisead}. Therefore, in this paper, the minimax method is proposed and introduced to solve the domain adaptation problem from a limited and more similar source domain to a rich and more diverse target domain. Moreover, in \cite{IWGAN}, the focus was on object recognition problems and the adversarial network was selected due to the advantages in classifying the objects. However, it is challenging for the state-of-the-art GAN approaches to realize the Nash equilibrium as compared to variance constrained autoencoders (VAEs) or Pixel RNNs \cite{Dualing}. Recent studies have shown that the autoencoder is advantageous at learning smooth latent state representations of the input data to reduce computational complexity, which has been exploited in speech enhancement \cite{ae}\cite{ae1}. Therefore, the importance-weighting (IW) method is exploited in the proposed techniques by using a variance constrained autoencoder (VCAE) to improve the speech enhancement performance.

The contributions of this paper are:

1) The importance of domain adaptation in unsupervised speech enhancement is confirmed and the IW method is proposed to utilize two classifiers with the variance constrained autoencoder to estimate the importance weights of the source samples. Besides, the improved performance is verified by the IEEE dataset \cite{IEEE}.

2) To strengthen the generalization performance of the domain adaptation method, the minimax method is proposed to transfer the model from a limited source domain to a rich target domain and the performance is confirmed with the VOICE BANK dataset \cite{VB}.

The organization of this paper is as follows: in Section II, two proposed methods, including the structure of the IW-VCAE approach, are shown in details. Section III presents the experimental settings, results, and discussions. Finally, the conclusions and future work are provided in Section IV. 
\section{PROPOSED METHOD}
In this section, the proposed methods and their comparisons to domain adaptation speech enhancement in different scenarios are presented. 
\begin{figure*}[htbp!]
\centering
\includegraphics[width=17.5cm, height=4cm]{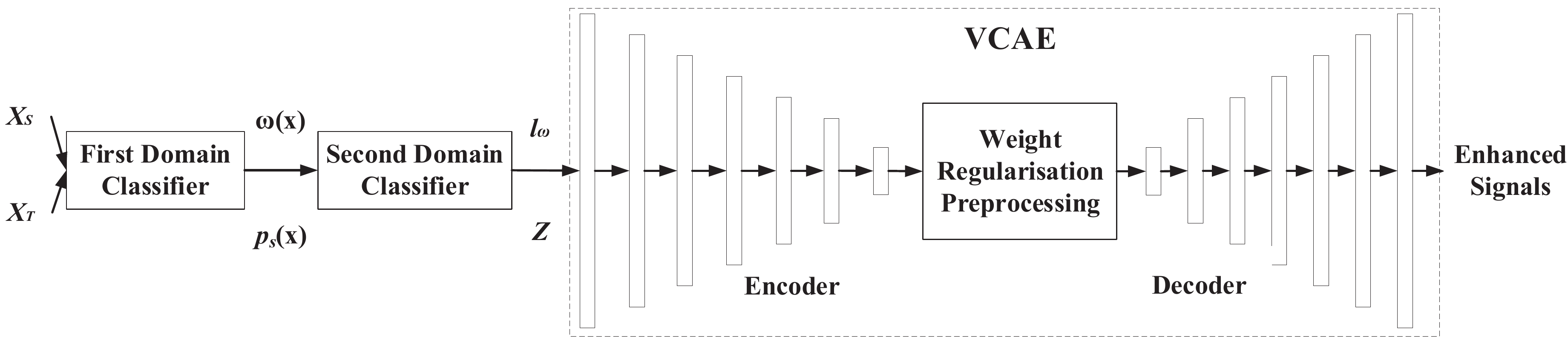}
\caption{The structure of the proposed importance weighting - variance constraint autoencoder (IW-VCAE). The features of the source and target domain mixtures, $\mathit{X_{S}}$ and $\mathit{X_{T}}$ respectively, are the inputs to the first classifier. The second classifier uses the weights and distributions of samples to estimate the loss function and the weighted features. Finally, the enhanced signals are obtained from the VCAE.}\centering
\end{figure*}
\subsection{Problem Statements and Domain Adaptation}
In the speech enhancement, the input and output spaces are denoted as $\mathit{X}$ and $\mathit{Y}$, respectively. We present the source domain as ($\mathit{X_{S}}$, $\mathit{Y_{S}}$, $\mathit{p_{S}}$) and refer to it with $\emph{S}$. Similarly, the target domain is denoted as ($\mathit{X_{T}}$, $\mathit{Y_{T}}$, $\mathit{p_{T}}$) and referred to by $\emph{T}$. The domain-specific functions are presented with the subscripts $\emph{S}$ and $\emph{T}$. For example, $p_{S}(\mathbf{x}, \mathbf{y})$ and $\mathit{p_{T}(\mathbf{x}, \mathbf{y})}$ are the source and target joint distributions, respectively. Moreover, $\mathit{p_{S}(\mathbf{x})}$ is for the source data marginal distribution and $\mathit{p_{T}(\mathbf{x}|\mathbf{y})}$ as the target class-conditional distribution.

Unsupervised domain adaptation (UDA) is a task to train a regression model on labeled data from a source domain to improve performance on a target domain, with access to only unlabeled data in the target domain \cite{adapt1}. In domain adaptation problems, according to the distribution comparisons between the source and the target domains, they are generally divided into two categories. The first is transferring from a rich and diverse domain to a limited and similar domain. In this case, the neural networks are trained by more weighted samples and classes. The second is transferring from a limited and similar domain to a rich and diverse domain and is much more challenging compared to the first category in speech enhancement. In order to perform the importance of the domain adaptation, TABLE I shows the speech enhancement performance comparison using the same dataset in the training stage but different datasets in the testing stage \cite{VCAE}.

From TABLE I, it can be observed that the speech enhancement performance is reduced in all scenarios compared to the same dataset, VOICE BANK, utilized in both the training and testing stages. The first reason for the generalization problem is the difference in the tones caused by the various microphones in datasets recording \cite{ae}. Furthermore, the environmental factors, including the distance of the speakers to the microphone, the location of the speakers, and the background interferences, play an important role in speech quality. Although the selected speakers of the dataset are independent, they are more likely selected from the same territorial area where the acoustic features such as the accent are highly similar. Therefore, the proposed methods address the domain adaptation problem by identifying the source samples that are potentially from the outlier weights and reducing the shift of shared weights between the source and target domains. The proposed IW-VCAE method is described in Section II. B.
\subsection{Importance-Weighting VCAE}
In neural network training, the weighting samples from the source domain to the target domain are exclusively exploited in the covariate shift due to the importance of generating better regression models. As in \cite{IW1}, the authors focused the source distribution to correct the probability of the target distribution, an importance-weighted classifier is required to learn and weight the source and the target samples. Therefore, a generalization error bound is added and the difference between the true target error of the classifier, $e_{T}(\mathbf{x})$, and the empirical weighted source error, $\hat{e}_{S}(\mathbf{x})$, at sample size, $n$, can be presented as \cite{TL}:
\\
\begin{small}
\begin{equation}
\frac{e_{T}(\mathbf{x})-\hat{e}_{S}(\mathbf{x})}{2}\leq\!\sqrt[5/4]{\mathrm{D}_{2}(\mathit{p_{T}}||\mathit{p_{S}})}\sqrt[3/8]{\frac{h}{n}\mathrm{log}(\frac{ne}{h})\!+\!\frac{1}{n}\mathrm{log}(\frac{4}{\delta })}
\end{equation}
\end{small}
\\
where the probability is at least 1-$\delta$, for 0 $<\delta \leqslant $ 1. Moreover, $\mathrm{D}_{2}(\mathit{p_{T}}||\mathit{p_{S}})$ represents the second-order R$\acute{\mathrm{e}}$nyi divergence which is related to R$\acute{\mathrm{e}}$nyi entropy as a measurement of information that satisfies almost the same axioms as Kullback-Leibler divergence (KLD) \cite{Renyi}. The second-order refers to two distributions as $\mathit{p_{T}}$ and $\mathit{p_{S}}$. The pseudo-dimension of the finite hypothesis space is represented as $h$, which is exploited to show the complexity of the hypothesis space \cite{pseudo}. $e$ is the Euler's number. Furthermore,  the weights are required to be nonzero, and $\mathrm{D}_{2}(\mathit{p_{T}}||\mathit{p_{S}})$, $n$ and $h$ are finite. In order to generalize domain adaptation performance, $\mathrm{D}_{2}(\mathit{p_{T}}||\mathit{p_{S}})$ is trained to maximum for the significantly different source and target domains, in which case the generalization difference range is varied.

In order to estimate the weights, the KLD between the true target distribution and the IW source distribution can be simplified as:
\begin{small}
\begin{equation}
\begin{aligned}
D_{KL}(\omega ,p_{S},p_{T})\!&=\! \int_{X}^{}p_{T}(\mathbf{x})\mathrm{log}(\frac{p_{T}(\mathbf{x})}{p_{S}(\mathbf{x})\omega(\mathbf{x})})\mathrm{d}\mathbf{x}\\
&=\!\int_{X}^{}p_{T}(\mathbf{x})\mathrm{log}(\frac{p_{T}(\mathbf{x})}{p_{S}(\mathbf{x})})\mathrm{d}\mathbf{x}\!-\!\int_{X}^{}\!p_{T}(\mathbf{x})\mathrm{log}(\omega(\mathbf{x}))\mathrm{d}\mathbf{x}
\end{aligned}
\end{equation}
\end{small}
where $\omega(\mathbf{x})$ represents the weight of the sample and $\mathbf{x}$ is the sample. In the right-hand side of the equation, the first term is independent of the weights. Therefore, the second term $\int_{X}^{}\!\mathit{p_{T}}(\mathbf{x})\mathrm{log}\omega(\mathbf{x})\mathrm{d}\mathbf{x}$ is utilized in the optimization function. Moreover, as the expected value of the logarithmic weights with the responding target domain distribution, the second term is estimated with the unlabeled target samples as:
\begin{equation}
\mathbb{E}_{T}(\mathrm{log}(\omega(\mathbf{x})))\approx \frac{1}{m}\sum_{j}^{m}\mathrm{log}(\omega(z_{j}))
\end{equation}
where $m$ represents the sample size of the target domain and the $z_{j}$ denotes the $j$th observation drawn from the target domain. Fig. 1. presents the overview of the proposed IW-VCAE method.

As shown in Fig. 1, in the training stage, the feature extraction for the source domain $X_{S}$ and the generated target domain $X_{T}$ are input to the first domain classifier to obtain the importance weights of the source samples.
\begin{equation}
C(\mathit{X})=\mathit{p}(y=1|\mathit{X})=\sigma (\mathit{X})
\end{equation}
where $C$ is the classifier and $\sigma(\cdot )$ is the logistic sigmoid function. $\mathit{X}$ is the sample in the features space after feature extraction. In the backward propagation training, the first domain classifier is converged to the optimal value based on the input and provides the sampling likelihood with the source distribution. Hence, the weights for source samples from outlier classes will be smaller than the shared class samples. In order to obtain the source sample importance weights, the samples are assumed to have relatively small weights as compared with the samples from the shared weights. Therefore, the weights can be defined and normalized as:
\begin{equation}
\omega(\mathbf{x})=\frac{1-C(\mathit{X})}{\mathbb{E}_{\mathbf{x}\sim \mathit{p}_{S}(\mathbf{x})}(1-C(\mathit{X}))}
\end{equation}
such that $\mathbb{E}_{\mathrm{x}\sim \mathit{p}_{S}(\mathrm{x})}\omega(\mathrm{x})=1$. It can be seen that if $\omega(\mathbf{x})$ is relatively small, $C(\mathit{X})$ is large and $\frac{\mathit{p}_{S}(\mathrm{x})}{\mathit{p}_{T}(\mathrm{x})}$ is small because $\omega(\mathbf{x})$ also represents density ratio between source and target features. Hence, the weights for source samples from outlier classes will be smaller than the shared class samples. However, in order to reduce the Jensen-Shannon divergence between the source and target densities, the second domain classifier is introduced based on the output of the $C$, namely $C_{2}$ \cite{Jensen}.

In order to reduce the domain shift, the importance weights are added to the source samples for the second domain classifier $C_{2}$ and the loss function can be presented as:
\begin{equation}
\begin{aligned}
\mathit{l_{\omega}}\left(C_{2}, X_{S}, X_{T}\right)=\mathbb{E}_{\mathbf{x} \sim p_{S}(\mathbf{x})} \omega(\mathbf{x}) \log \left(C_{2}\left(X_{S}\right)\right) \\
+\mathbb{E}_{\mathbf{x} \sim p_{T}(\mathbf{x})} (1-\log \left(C_{2}\left(X_{T}\right)\right))
\end{aligned}
\end{equation}
where $\omega(\mathbf{x})$ is a function of the first domain classifier and independent of $C_{2}$. Because $\omega(\mathbf{x})$ is normalized, $\omega(\mathbf{x})\mathit{p}_{S}(\mathbf{x})$ can be regarded as a probability density function:
\begin{equation}
\mathbb{E}_{\mathbf{x}\sim \mathit{p}_{S}(\mathbf{x})}\omega(\mathbf{x})=\int \omega(\mathbf{x})\mathit{p}_{S}(\mathbf{x})\mathrm{d}\mathbf{x}=1
\end{equation}
To obtain the optimal value of $C_{2}$, the loss function can be reformulated as:
\begin{equation}
\begin{aligned}
\mathit{l_{\omega }}(X_{T})&=\int_{x}^{}\omega(\mathbf{x})\mathit{p}_{S}(\mathbf{x})\mathrm{log}(\frac{\omega(\mathbf{x})\mathit{p}_{S}(\mathbf{x})}{\omega(\mathbf{x})\mathit{p}_{S}(\mathbf{x})+\mathit{p}_{T}(\mathbf{x})})\\
&\quad +\mathit{p}_{T}(\mathbf{x})\mathrm{log}(\frac{\mathit{p}_{T}(\mathbf{x})}{\omega(\mathbf{x})\mathit{p}_{S}(\mathbf{x})+\mathit{p}_{T}(\mathbf{x})})\\
\end{aligned}
\end{equation}
Besides, the loss function can be simplified as:
\begin{equation}
\mathit{l_{\omega }}(X_{T})=2 \mathit{JS}[\omega(\mathbf{x})\mathit{p}_{S}(\mathbf{x})||\mathit{p}_{T}(\mathbf{x})]
\end{equation}
where $\mathit{JS}$ is the Jensen-Shannon divergence between the weighted source and target densities based on the feature extractor and is optimized as $\omega(\mathrm{x})\mathit{p}_{S}(\mathrm{x})=\mathit{p}_{T}(\mathrm{x})$. Furthermore, after $\mathit{JS}$ is reduced by the weighted samples domain adaptation, the VCAE is utilized to improve the speech enhancement performance. The pseudo code of the proposed IW-VCAE method is summarized as Algorithm 1.


\algblockdefx[noEndBlock]{noEnd}{noEndEnd}
[1]{#1}
{\vspace{-\baselineskip}}


\begin{algorithm}

  
  \SetKwInOut{Input}{input}\SetKwInOut{Output}{output}

  \Input{Extracted features $\mathit{X_{S}}$ and $\mathit{X_{T}}$ }
  \Output{Desired signal}
  \BlankLine
  Initialize the feature extractors
  \For{$epoch\leftarrow 1, 2, ...,$  $30$}{
    \While{$\mathbb{E}_{\mathrm{x}\sim \mathit{p}_{S}(\mathrm{x})}\omega(\mathrm{x})=1$}{
      Train $C(\mathit{X})$ by as Eq. (4) in Section II. B\;
    }
    \While{$\omega(\mathrm{x})$ is normalized}{
      Train $C_{2}$ by minimizing the loss function as Eq. (9) in Section II. B\;
    }
    Constrain $X_{T}$ by minimizing the entropy\;
    Sample $\mathbf{x}_{i}$ with $\mathbf{\omega}_{i}$ from the classifier\;
    Sample $\mathbf{z}_{i}$ from the classifier and prior p(z)\;
    Compute the gradients of hyperparameters $\theta$ and $\lambda$ \
    \eIf{$\mathbf{x}_{i}$ or $\mathbf{z}_{i}$ is overfitting}{
      Compute and normalize the latent vector\;
    }{
      Train the VCAE by optimizing Eq. (10) in Section II. B\;
    } }
  \caption{IW-VCAE pseudo code.}\label{algo_disjdecomp}
\end{algorithm}

In the training stage, given by the underlying features $Z$, the likelihood of the domain samples is maximized as \cite{VCAE}:
\begin{equation}
\begin{aligned}
\underset{\phi,\theta }{\mathrm{max}}\ &\mathbb{E}_{X_{S}\sim \mathit{p}(\mathbf{x})}\mathbb{E}_{Z\sim \mathit{p}_{\phi}(\mathbf{z}|\mathbf{x})}\left \{\mathrm{log}[\mathit{p}_{\theta}(X_{S}|Z)]\right \}\\
-&\lambda \left |\mathbb{E}_{Z\sim \mathit{p}_{\phi}(\mathbf{z})}[\left \|Z- \mathbb{E}_{Z\sim \mathit{p}_{\phi}(\mathbf{z})} [Z] \right \|_{2}^{2}]-v \right |
\end{aligned}
\end{equation}
where $\mathit{p}_{\phi}(\cdot )$ and $\mathit{p}_{\theta}(\cdot )$ represent the encoder and decoder distributions with the parameters, $\phi$ and $\theta$, in the network, respectively \cite{VCAE}. Moreover, $\lambda$ is a hyperparameter, $\mathbf{z}$ is the latent feature, and $v$ is the desired summed variance of the distribution. After the likelihood is optimized in the network, less desired signals are obtained at the terminals of the input block window than in the middle as the same mixture and the desired speech block sizes. Therefore, additional weighted samples are input to the encoder and provide information about the signal performance at the window boundaries. The extracted features of 1000 importance weighted noisy samples are utilized as the input to enhance the desired signal. Additionally, a weight regularization preprocessing block is added between the encoder and the decoder to address the overfitting problem caused in the training stage \cite{WR}. The VCAE is penalized based on the size of the network weights in the training stage.
\subsection{Minimax}
Under the specific settings, the IW method can significantly improve speech enhancement performance. However, in some further challenging scenarios in which the assumptions including limited source samples, the domain adaptation is possible to be detrimental to the performance. Furthermore, well-trained models can encounter different sorts of settings in real-world scenarios. Therefore, minimax is proposed to train the classifier with the worst-case weights. 

A risk-minimization model is generalized if the decisions made by the information on one specific problem are available for the other similar problems. Initially, in order to guarantee the improvement, the worst-case setting is assumed, which is formalized as the minimax optimization method. The main concept of the proposed method is that the risk is minimized using the classifier parameters and the strengthening variables are maximized. Because the IW classifier is sensitive to poorly trained weights, the risk is minimized using the worst-case weights:
\begin{equation}
\underset{h\epsilon \emph{H}}{min}\ \underset{\omega \epsilon \emph{H} }{max}\ \frac{1}{n}\sum_{i=1}^{n}\emph{l}(h(\mathbf{x}_{i}),z_{j})\omega _{i}
\end{equation}
where $h(\mathbf{x}_{i})$ represents the decision made by the classifier and $\emph{H}$ is the finite hypothesis space. The pseudo code of the proposed minimax method is summarized as Algorithm 2.

\begin{algorithm}
  \SetKwInOut{Input}{input}\SetKwInOut{Output}{output}

  \Input{Extracted features $\mathit{X_{S}}$ and $\mathit{X_{T}}$ with worst-case $\omega$}
  \Output{Desired signal}
  \BlankLine
  \For{$epoch\leftarrow 1, 2, ...,$  $30$}{
    \eIf{$\omega _{i}\leqslant 0$ or $\omega _{i}> 1$}{
      Add a vector norm penalty to the Robust Bias-Aware classifier\;
    }{
      Train Robust Bias-Aware classifier by optimizing Eq. (11) in Section II. C\;
    }
    Sample $\mathbf{x}_{i}$ with $\mathbf{\omega}_{i}$ from the classifier\;
    Sample $\mathbf{z}_{i}$ from the classifier and prior p(z)\;
    Compute the gradients of hyperparameters $\theta$ and $\lambda$ \
    \eIf{$\mathbf{x}_{i}$ or $\mathbf{z}_{i}$ is overfitting}{
      Compute and normalize the latent vector\;
    }{
      Train the VCAE by optimizing Eq. (10) in Section II. B\;
    }
    }
  \caption{Minimax pseudo code.}\label{algo_disjdecomp}
\end{algorithm}

The estimated weights are constrained as:
\begin{equation}
0 < \omega _{i} \leqslant 1
\end{equation}
\begin{equation}
\left | \frac{1}{n}\sum_{i}^{n}\omega _{i}-1 \right |\leq \epsilon 
\end{equation}
where $\left |\cdot \right |$ is the absolute value operator and $\epsilon$ is a small value variable to ensure that the estimated weights are constrained to 1 and the constraints match the non-parametric weight estimators. Different from the proposed IW method utilizing two classifiers, the minimax approach uses only one as Robust Bias-Aware classifier. The conditional label distribution is provided and is robust to the worst-case logarithmic loss for the target domain distribution while matching feature expectation constraints from the source domain distribution \cite{KB}.

From Algorithm 1 and Eq. (5), in the training stage, $\mathbb{E}_{\mathrm{x}\sim \mathit{p}_{S}(\mathrm{x})}\omega(\mathrm{x})$ is converged to 1. Thus, the density ratio between the source and target features is normalized and the shift of shared weights between the source and target domains is reduced. However, the proposed minimax method uses the worst-case weights for the initialization at the input of the classifier and the weights are constrained to reduce the overfitting problem during the training.  Moreover, the minimax method requires only one classifier and reduces computational complexity. Thus, the minimax method is better suitable for the real-world scenarios where transferring from a small dataset containing limited sorts of samples to a large dataset with rich samples. 

\section{EXPERIMENTAL RESULTS}
\begin{table*}[htbp!]
\caption{SPEECH ENHANCEMENT PERFORMANCE COMPARISON IN TERMS OF {\bfseries \upshape{STOI (IN $\%$)}} WITH DIFFERENT TRAINING METHODS, SNR LEVELS AND NOISES. THE {\bfseries VOICE BANK} DATASET IS USED IN THE TRAINING STAGE AND {\bfseries TIMIT} DATASET IS FOR THE TESTING STAGE. {\bfseries BOLD} INDICATES THE BEST RESULTS. \textit{Italic} SHOWS THE PROPOSED METHODS. EACH RESULT IS AVERAGE OF 900 EXPERIMENTS.}
\centering
\small\addtolength{\tabcolsep}{-3pt}
\begin{tabular}{|c|c|c|c|c|c|c|c|c|c|c|}
\hline
Noise  & \multicolumn{3}{|c|}{psquare} & \multicolumn{3}{|c|}{living} & \multicolumn{3}{|c|}{station}&Average\\
\hline
SNR level (dB) & -5  & 0 & 5 & -5  & 0  & 5 & -5  & 0  & 5 &  \\     
 \hline
 Unprocessed & 58.5 & 61.3& 66.6& 57.3& 59.2& 63.8& 56.8&59.7& 60.0& 60.1 \\
 \hline
SEGAN \cite{Joint}  &70.0 & 73.6 &78.9&68.7 & 70.1 & 73.5&65.3 & 66.8 & 69.9&70.7 \\
 \hline
VCAE \cite{VCAE}  &73.4 & 76.9 & 82.1 &70.5 & 72.6 & 77.0&69.4 & 71.3 & 74.9&74.2 \\
\hline
DANN \cite{DANN} &73.8 &78.1 &83.0 &71.2 &74.0 &79.1 &70.7 &71.9 &75.3 &75.2\\
\hline
IWAE \cite{IWA} &74.1 &78.2 &83.0 &71.6 &74.2 &79.2 &70.5 &72.1 &75.2 &75.3\\
\hline
ADDA \cite{ADDA} &74.2 &78.4 &83.2 &71.5 &74.2 &79.6 &70.9 &72.3 &75.5 &75.5\\
\hline
\textit{Importance-weighting}   & 74.9 & 79.7  & 84.8&72.1 & 74.9 & 80.6&71.5 & 73.2 & 77.1&76.5\\
 \hline
\textit{Minimax}   &{\bfseries 75.3} &{\bfseries 79.9}  &{\bfseries 84.9}&{\bfseries 73.0}&{\bfseries 75.7}&{\bfseries 81.6}&{\bfseries 73.2}&{\bfseries 74.6}&{\bfseries 78.1} &{\bfseries 77.4}\\
 \hline 
\end{tabular}
\end{table*}

\begin{table*}[htbp!]
\caption{SPEECH ENHANCEMENT PERFORMANCE COMPARISON IN TERMS OF {\bfseries PESQ} WITH DIFFERENT TRAINING METHODS, SNR LEVELS AND NOISES. THE {\bfseries VOICE BANK} DATASET IS USED IN THE TRAINING STAGE AND {\bfseries TIMIT} DATASET IS FOR THE TESTING STAGE. {\bfseries BOLD} INDICATES THE BEST RESULTS. \textit{Italic} SHOWS THE PROPOSED METHODS. EACH RESULT IS AVERAGE OF 900 EXPERIMENTS.}
\centering
\small\addtolength{\tabcolsep}{-3pt}
\begin{tabular}{|c|c|c|c|c|c|c|c|c|c|c|}
\hline
Noise  & \multicolumn{3}{|c|}{psquare} & \multicolumn{3}{|c|}{living} & \multicolumn{3}{|c|}{station}&Average\\
\hline
SNR level (dB) & -5  & 0 & 5 & -5  & 0  & 5 & -5  & 0  & 5 &  \\     
 \hline
 Unprocessed & 1.59 & 1.70& 1.81& 1.50& 1.57& 1.67& 1.49&1.55& 1.66& 1.61 \\
 \hline
SEGAN \cite{Joint}  &1.79 & 1.92 & 2.08&1.72 & 1.80 & 1.94&1.71 & 1.74 & 1.90&1.86 \\
 \hline
VCAE \cite{VCAE}  &1.84 & 2.00 & 2.27&1.77 & 1.84 & 1.99&1.75 & 1.79 & 1.95&1.91 \\
\hline
DANN \cite{DANN} &1.91 &2.11 &2.39 &1.79 &1.93 &2.09 &1.77 &1.88 &2.05 &1.99\\
 \hline
 IWAE \cite{IWA} &1.95 &2.11 &2.36 &1.80 &1.93 &2.05 &1.81 &1.90 &2.02 &1.99\\
 \hline
 ADDA \cite{ADDA} &1.94 &2.13 &2.40 &1.81 &1.95 &2.10 &1.79 &1.91 &2.07 &2.01\\
 \hline
\textit{Importance-weighting}    &2.00 & 2.18 & 2.46&1.85 & 2.02 & 2.23&1.83 & 2.00 & 2.18&2.09\\
 \hline
\textit{Minimax}   &{\bfseries 2.04} &{\bfseries 2.21}  &{\bfseries 2.50}&{\bfseries 1.89}&{\bfseries 2.09}&{\bfseries 2.31}&{\bfseries 1.88}&{\bfseries 2.07}&{\bfseries 2.28} &{\bfseries 2.15}\\
 \hline 
\end{tabular}
\end{table*}

\begin{table*}[htbp!]
\caption{SPEECH ENHANCEMENT PERFORMANCE COMPARISON IN TERMS OF {\bfseries \upshape{fwSNRseg (dB)}} WITH DIFFERENT TRAINING METHODS, SNR LEVELS AND NOISES. THE {\bfseries VOICE BANK} DATASET IS USED IN THE TRAINING STAGE AND {\bfseries TIMIT} DATASET IS FOR THE TESTING STAGE. {\bfseries BOLD} INDICATES THE BEST RESULTS. \textit{Italic} SHOWS THE PROPOSED METHODS. EACH RESULT IS AVERAGE OF 900 EXPERIMENTS.}
\centering
\small\addtolength{\tabcolsep}{-3pt}
\begin{tabular}{|c|c|c|c|c|c|c|c|c|c|c|}
\hline
Noise  & \multicolumn{3}{|c|}{psquare} & \multicolumn{3}{|c|}{living} & \multicolumn{3}{|c|}{station}&Average\\
\hline
SNR level (dB) & -5  & 0 & 5 & -5  & 0  & 5 & -5  & 0  & 5 &  \\     
 \hline
 Unprocessed & 3.37 & 3.98& 4.81& 3.26& 3.69& 4.28& 3.04&3.41& 3.93& 3.75 \\
 \hline
SEGAN \cite{Joint}  &8.90 & 9.66 & 10.23&8.65 & 9.05 & 9.89&8.31 & 8.92 & 9.35&9.22 \\
 \hline
VCAE \cite{VCAE}  &9.64 & 10.52 & 11.83&9.28 & 10.11 & 10.80&8.86 & 9.55 & 10.38&10.10 \\
 \hline
DANN \cite{DANN} &10.97 &12.11 &13.08 &10.42 &11.06 &12.04 &9.99 &11.17 &11.73 &11.39\\
\hline
IWAE \cite{IWA} &11.17 &12.20 &13.01 &10.66 &11.25 &11.99 &10.14 &11.28 &11.63 &11.48\\
\hline
ADDA \cite{ADDA} &11.34 &12.27 &13.21 &10.79 &11.40 &12.15 &10.27 &11.33 &11.86 &11.62\\
\hline
\textit{Importance-weighting}    &12.06 & 13.84 & 15.60&11.77 & 12.34 & 13.24&11.59 & 12.05 & 12.97&12.84\\
 \hline
\textit{Minimax}   &{\bfseries 12.44} &{\bfseries 14.27}  &{\bfseries 16.12}&{\bfseries 12.01}&{\bfseries 13.09}&{\bfseries 14.33}&{\bfseries 11.74}&{\bfseries 12.38}&{\bfseries 14.09} &{\bfseries 13.40}\\
 \hline 
\end{tabular}
\end{table*}
In this section, the proposed methods are evaluated by various datasets and compared to the state-of-the-art methods via intelligibility metrics.
\subsection{Datasets and Network Parameters}
In order to evaluate the speech enhancement and domain adaptation performance, in the training stage, 120 clean utterances of 20 speakers from the VOICE BANK dataset \cite{VB} and 600 clean utterances of 60 speakers from the IEEE dataset \cite{IEEE} are randomly selected in two subsections of the experiments, respectively. The VOICE BANK dataset already constitutes the largest corpora of British English and the IEEE dataset contains speech data of American English speakers. However, in the testing stage, 900 clean utterances of 90 speakers from the TIMIT dataset \cite{TIMIT} are utilized to evaluate the performance, which are common for the two subsections. Besides, 60 clean utterances of eight major dialects of American English are randomly selected from the TIMIT dataset to generate the validation dataset. Three non-speech noise interferences are mixed with clean speech utterances, and the noises are $psquare$, $living$, and $station$. In the testing stage, the noise interferences are seen at the training stage. Therefore, the trained neural network is able to distinguish noises from the target speech signals. The noise interferences are selected from the Demand database  \cite{Demand} for our evaluations. Each noise scene has a unique example and four minutes long, and it is divided into two clips with an equal length. One is used to match the lengths of the speech signals to generate training data and another is used to generate validation and testing data \cite{ICS}. Hence, in total, there are 1080 mixtures (120$\times$3$\times$3) and 5400 mixtures (600$\times$3$\times$3) for the VOICE BANK and the IEEE in the training data, respectively. Furthermore, there are 540 mixtures (60$\times$3$\times$3) in the validation data, 8100 mixtures (900$\times$3$\times$3) in the testing data based on three SNR levels (-5 dB, 0 dB, and 5 dB).

In this study, amplitude modulation spectrogram (AMS) \cite{AMS2}, relative spectral transform perceptual linear prediction (RASTA-PLP) \cite{PLP2}, and delta-spectral cepstral coefficients (DSCC) \cite{DSCC} are exploited as the features. AMS is extensively used in speech processing due to outstanding performance. RASTA-PLP is a linear prediction feature and suitable for processing the temporal dynamics of speech. Although proposed for speech recognition, DSCC performs temporal differencing in the spectra and is applied in speech enhancement.

Besides, both the baselines and proposed methods are trained by using the RMSprop algorithm with a learning rate of 0.001 \cite{Student}. The number of epochs is 30, and the batch size is 512. As for the proposed methods, the input and output block sizes are 62.5 and 37.5 milliseconds that allow 1000 noisy samples exploited as input for the central 600 samples. Additionally, seven 1D-convolutional layers with 64, 64, 128, 128, 256, 256, and 512 filters and 31 kernels are composed in the encoder. The Leaky-ReLU ($\alpha $ = 0.1) activation is utilized in the first six layers while the last layer uses a linear activation. The strides of the middle five layers are set to two, however, the first and last layers have strides of one. In the final convolutional layer, the output is obtained by a dense linear-layer with 660 neurons. In the decoder, seven 1D-convolutional layers with 512, 256, 256, 128, 128, 64, and 64 filters are applied with 31 kernels in each layer. We define $\lambda$ = 0.01, $\phi$ = $1\times 10^{-6}$, and $v$ = 330.
\subsection{Comparisons and Performance Measurements}
In \cite{VCAE}, the original VCAE method has been confirmed to outperform the SE-WaveNet \cite{sewav}. Therefore, the proposed methods are compared to the SEGAN \cite{Joint} and the original VCAE \cite{VCAE} approaches because these are recent state-of-the-art methods in speech enhancement. Additionally, the phase information is not utilized in the proposed methods to keep the computational complexity because of the trade-off between the computational cost and the enhancement performance \cite{Student}.
\begin{table*}[htbp!]
\caption{SPEECH ENHANCEMENT PERFORMANCE COMPARISON IN TERMS OF {\bfseries \upshape{STOI (IN $\%$)}} WITH DIFFERENT TRAINING METHODS, SNR LEVELS AND NOISES. THE {\bfseries IEEE} DATASET IS USED IN THE TRAINING STAGE AND {\bfseries TIMIT} DATASET IS FOR THE TESTING STAGE. {\bfseries BOLD} INDICATES THE BEST RESULTS. \textit{Italic} SHOWS THE PROPOSED METHODS. EACH RESULT IS AVERAGE OF 900 EXPERIMENTS.}
\centering
\small\addtolength{\tabcolsep}{-3pt}
\begin{tabular}{|c|c|c|c|c|c|c|c|c|c|c|}
\hline
Noise  & \multicolumn{3}{|c|}{psquare} & \multicolumn{3}{|c|}{living} & \multicolumn{3}{|c|}{station}&Average\\
\hline
SNR level (dB) & -5  & 0 & 5 & -5  & 0  & 5 & -5  & 0  & 5 &  \\     
 \hline
 Unprocessed & 58.5 & 61.3& 66.6& 57.3& 59.2& 63.8& 56.8&59.7& 60.0& 60.1 \\
 \hline
SEGAN \cite{Joint}  &72.0 & 75.4 &80.7&69.9 & 72.2 & 76.8&68.1 & 68.4 & 73.7&73.0 \\
 \hline
VCAE \cite{VCAE}  &73.8 & 77.4 & 82.6 &71.2 & 73.0 & 77.9&70.3 & 72.0 & 75.8&75.0 \\
 \hline
IWAE \cite{IWA}  &74.6 & 78.1 & 82.5 &72.1 & 74.2 & 77.7&71.4 & 72.2 & 75.6&75.4 \\
 \hline
ADDA \cite{ADDA} &74.8 &78.7 &83.4 &72.1 &74.6 &79.6 &71.7 &72.7 &76.1 &75.9\\
\hline 
\textit{Minimax}   & 75.4 & 80.0  & 85.1&73.3 & 76.1 & 82.1&73.2 & 74.7 & 78.3&77.5\\
 \hline
\textit{Importance-weighting}   &{\bfseries 75.6} &{\bfseries 80.3}  &{\bfseries 85.6}&{\bfseries 73.4}&{\bfseries 76.2}&{\bfseries 82.4}&{\bfseries 73.4}&{\bfseries 74.9}&{\bfseries 78.7} &{\bfseries 77.9}\\
 \hline 
\end{tabular}
\end{table*}

\begin{table*}[htbp!]
\caption{SPEECH ENHANCEMENT PERFORMANCE COMPARISON IN TERMS OF {\bfseries PESQ} WITH DIFFERENT TRAINING METHODS, SNR LEVELS AND NOISES. THE {\bfseries IEEE} DATASET IS USED IN THE TRAINING STAGE AND {\bfseries TIMIT} DATASET IS FOR THE TESTING STAGE. {\bfseries BOLD} INDICATES THE BEST RESULTS. \textit{Italic} SHOWS THE PROPOSED METHODS. EACH RESULT IS AVERAGE OF 900 EXPERIMENTS.}
\centering
\small\addtolength{\tabcolsep}{-3pt}
\begin{tabular}{|c|c|c|c|c|c|c|c|c|c|c|}
\hline
Noise  & \multicolumn{3}{|c|}{psquare} & \multicolumn{3}{|c|}{living} & \multicolumn{3}{|c|}{station}&Average\\
\hline
SNR level (dB) & -5  & 0 & 5 & -5  & 0  & 5 & -5  & 0  & 5 &  \\     
 \hline
 Unprocessed & 1.59 & 1.70& 1.81& 1.50& 1.57& 1.67& 1.49&1.55& 1.66& 1.61 \\
 \hline
SEGAN \cite{Joint}  &1.85 & 2.06 & 2.32&1.74 & 1.85 & 2.11&1.76 & 1.80 & 2.00&1.94 \\
 \hline
VCAE \cite{VCAE}  &1.89 & 2.11 & 2.40&1.79 & 1.88 & 2.14&1.79 & 1.85 & 2.04&2.00 \\
\hline
DANN \cite{DANN} &1.96 &2.19 &2.58 &1.86 &2.00 &2.28 &1.85 &1.96 &2.11 &2.09\\
 \hline
 IWAE \cite{IWA} &2.00 &2.19 &2.55 &1.91 &2.02 &2.25 &1.87 &1.97 &2.08 &2.09\\
 \hline
 ADDA \cite{ADDA} &2.01 &2.23 &2.60 &1.90 &2.05 &2.29 &1.88 &2.00 &2.12 &2.12\\
 \hline
\textit{Minimax}    &2.09 & 2.28 & 2.63&1.92 & 2.13 & 2.40&1.89 & 2.10 & 2.36&2.20\\
 \hline
\textit{Importance-weighting}   &{\bfseries 2.11} &{\bfseries 2.30}  &{\bfseries 2.67}&{\bfseries 1.93}&{\bfseries 2.14}&{\bfseries 2.42}&{\bfseries 1.89}&{\bfseries 2.11}&{\bfseries 2.38} &{\bfseries 2.22}\\
 \hline 
\end{tabular}
\end{table*}

\begin{table*}[htbp!]
\caption{SPEECH ENHANCEMENT PERFORMANCE COMPARISON IN TERMS OF {\bfseries \upshape{fwSNRseg (dB)}} WITH DIFFERENT TRAINING METHODS, SNR LEVELS AND NOISES. THE {\bfseries IEEE} DATASET IS USED IN THE TRAINING STAGE AND {\bfseries TIMIT} DATASET IS FOR THE TESTING STAGE. {\bfseries BOLD} INDICATES THE BEST RESULTS. \textit{Italic} SHOWS THE PROPOSED METHODS. EACH RESULT IS AVERAGE OF 900 EXPERIMENTS.}
\centering
\small\addtolength{\tabcolsep}{-3pt}
\begin{tabular}{|c|c|c|c|c|c|c|c|c|c|c|}
\hline
Noise  & \multicolumn{3}{|c|}{psquare} & \multicolumn{3}{|c|}{living} & \multicolumn{3}{|c|}{station}&Average\\
\hline
SNR level (dB) & -5  & 0 & 5 & -5  & 0  & 5 & -5  & 0  & 5 &  \\     
 \hline
 Unprocessed & 3.37 & 3.98& 4.81& 3.26& 3.69& 4.28& 3.04&3.41& 3.93& 3.75 \\
 \hline
SEGAN \cite{Joint}  &9.69 & 10.03 & 10.47&9.00 & 9.35 & 10.08&8.86 & 9.40 & 9.77&9.64 \\
 \hline
VCAE \cite{VCAE}  &10.00 & 10.89 & 11.33&9.70 & 10.42 & 10.97&9.55 & 9.98 & 10.12&10.33 \\
 \hline
 DANN \cite{DANN} &10.88 &12.04 &15.23 &10.77 &12.02 &13.90 &10.78 &11.39 &13.51 &11.05\\
\hline
 IWAE \cite{IWA} &11.25 &12.31 &15.16 &11.06 &12.17 &13.76 &10.99 &11.48 &13.50 &12.40\\
\hline
 ADDA \cite{ADDA} &11.46 &12.57 &15.40 &11.11 &12.25 &14.03 &11.09 &11.52 &13.61 &12.56\\
\hline
\textit{Minimax}    &12.50 & 14.36 & 16.20&12.10 & 13.21 & 14.48&11.89 & 12.49 & 14.34&13.51\\
 \hline
\textit{Importance-weighting}   &{\bfseries 14.37} &{\bfseries 16.19}  &{\bfseries 18.17}&{\bfseries 14.26}&{\bfseries 15.14}&{\bfseries 16.69}&{\bfseries 13.95}&{\bfseries 14.41}&{\bfseries 16.44} &{\bfseries 15.52}\\
 \hline 
\end{tabular}
\end{table*}
In the SEGAN setup, the generative network, G, consisted of 22 1D-strided convolutional layers of 31 filters and 2 strides as well as the discriminative network, D \cite{Joint}. The resulting dimensions in each layer, being samples$\times$feature maps, are 16384$\times$1, 8192$\times$16, 4096$\times$32, 2048$\times$32, 1024$\times$64, 512$\times$64, 256$\times$128, 128$\times$128, 64$\times$256, 32$\times$256, 16$\times$512, and 8$\times$1024. As for the original VCAE method, five 1D-convolutional layers with 32, 32, 64, 128, and 128 filters and 31 kernels are composed in the encoder \cite{VCAE}. The Leaky-ReLU ($\alpha $ = 0.1) activation is utilized in the first four layers while the last layer uses a linear activation. The strides of the middle three layers are set to two, however, the first and last layers only have one stride each. The output from the final convolutional layer is processed by a dense linear-layer with 330 output neurons \cite{VCAE}. In the decoder, following a dense linear-layer with 75 $\times$ 128 output neurons, five 1D-convolutional layers with 64, 32, 16, 16, and one filters and 31 kernels are applied.

There are three intelligibility metrics, the short-time objective intelligibility (STOI), perceptual evaluation of speech quality (PESQ), and frequency-weighted segmental signal-to-noise ratio (fwSNRseg). The values of the STOI indicate the human speech intelligibility scores and are bounded in the range of [0, 1] \cite{LSTM4}. The PESQ refers to human speech quality scores and is bounded in the range of [-0.5, 4.5]. The fwSNRseg is calculated by computing the segmental signal-to-noise ratios (SNRs) in each spectral band and summing the weighed SNRs from all bands \cite{fw}. Higher values of these measurements imply that the desired speech signal is better extracted. Furthermore, in order to provide the level of improvement, the p-value of the t-test is calculated and the null hypothesis, $H_{0}$, is introduced to determine the level of statistical significance \cite{pvalue}. A p-value less than 0.05 (typically $\leq$  0.05) is statistically significant and indicates strong evidence against the null hypothesis.
\subsection{Results and Discussions}
As aforementioned, the experiments are divided into two subsections that one is to evaluate the model trained by the VOICE BANK dataset and the other is by the IEEE dataset. Both are tested by the TIMIT dataset. The p-value of the t-test and the spectra of different stages are presented in TABLE VIII and Fig. 2, respectively.
\begin{table*}[htbp!]
\caption{THE P-VALUE OF THE T-TEST AT 5$\%$ SIGNIFICANT LEVEL, COMPARISON OF THE PROPOSED METHODS WITH THE STATE-OF-THE-ART METHODS. $H_{0}$ DENOTES THE NULL HYPOTHESIS, AND (+) INDICATES THE IMPROVEMENT OF THE PROPOSED METHOD IS STATISTICALLY SIGNIFICANT AT THE 95$\%$ CONFIDENCE LEVEL. \textit{Italic} SHOWS THE PROPOSED METHODS}
\centering
\small\addtolength{\tabcolsep}{-3pt}
\begin{tabular}{|c|c|c|c|c|c|c|}
\hline
  & \multicolumn{2}{|c|}{STOI} & \multicolumn{2}{|c|}{PESQ} & \multicolumn{2}{|c|}{fwSNRseg}\\
\hline
 &p-value  & $H_{0}$ & p-value &$H_{0}$  & p-value  & $H_{0}$   \\     
 \hline
SEGAN-\textit{IW}  &1.01E-07 & (+) & 4.00E-06&(+) & 5.23E-08 & (+) \\
 \hline
VCAE-\textit{IW}  &2.91E-07 &(+) & 3.62E-06&(+) & 1.97E-08 & (+)  \\
 \hline
SEGAN-\textit{Minimax}    &1.62E-07 & (+) & 1.40E-06&(+) & 4.65E-08 & (+)\\
 \hline
VCAE--\textit{Minimax}  &2.91E-07 & (+) & 2.60E-06&(+) & 1.00E-07 & (+) \\
 \hline 
DANN--\textit{IW}  &1.58E-05 & (+) & 2.74E-05&(+) & 4.30E-06 & (+) \\
 \hline 
ADDA--\textit{Minimax}  &1.36E-04 & (+) & 2.96E-05&(+) & 7.48E-06 & (+) \\
 \hline 
\end{tabular}
\end{table*}
\subsubsection{Transferring from VOICE BANK to TIMIT}
In these experiments, the STOI, PESQ, and fwSNRseg performance of different methods using the VOICE BANK and the TIMIT corpora with different noises and SNR levels are shown in Tables II-IV.

From Tables II-IV, it is observed that the STOI, PESQ, and fwSNRseg performances are refined by the proposed methods compared to the state-of-the-art methods in all SNR levels and scenarios. On the one hand, the proposed methods reduce the shift of shared weights between the source and target domains. The source domain samples are weighted with two domain classifiers and the outlier samples are ignored as only a subset of weights involved in the target domain. On the other hand, the minimax method better performs than the IW method due to the significant difference between the source and target domains. Compared to the original VCAE method in Table I, speech enhancement performance is improved in all scenarios. For instance, for the $living$ noise, the proposed minimax method can achieve 75.7 over STOI (in $\%$) in 0 dB SNR level. However, the original VCAE method only achieves 73.6, although it is tested by the same dataset with the training stage. The weight regularization preprocessing block between the encoder and the decoder plays an important role in addressing the overfitting problem caused in the training stage. Furthermore, the proposed methods optimize the network structure on the number of layers, filter sizes, and feature extractors to further improve the speech enhancement.
\subsubsection{Transferring from IEEE to TIMIT}
In these experiments, the STOI, PESQ, and fwSNRseg performance of different methods using the IEEE and the TIMIT corpora with different noises and SNR levels are shown in Tables V-VII.

In overall evaluations, it is clear that the proposed methods outperform the state-of-the-art methods in all SNR levels and scenarios. However, the IW method performs better than the minimax method, which is different from Section III-C-1. The reason is that in Section III-C-2 of experiments, 600 clean utterances of 60 speakers from the IEEE dataset are randomly selected as the training set that is much richer than Section III-C-1, only 120 clean utterances of 20 speakers from the VOICE BANK dataset. Compared to the minimax method using the risk-minimization model to train the classifier under the worst-case weights, the IW method classifies the samples from the source domain outlier weights and is more applicable for general domain adaptation cases. 
\subsubsection{T-test and spectra}
In order to determine the level of statistical significance, the p-value of the t-test of STOI, PESQ, and fwSNRseg performance of pairs of different methods using the VOICE BANK and the IEEE corpora with different noises and SNR levels are shown in Table VIII. 

From Table VIII, in all comparisons between the proposed methods and the baselines, the p-values are less than 0.05 that indicates the statistically significant improvement of both the proposed methods. In the comparisons of the proposed methods and the baselines, the fwSNRseg performance is significantly improved. Moreover, the spectra of the clean speech signal, the mixture, and the estimated signals from the baselines and the proposed methods are presented in Fig. 2.

\begin{figure}[h!]
\centering
\includegraphics[width=9cm, height=7cm]{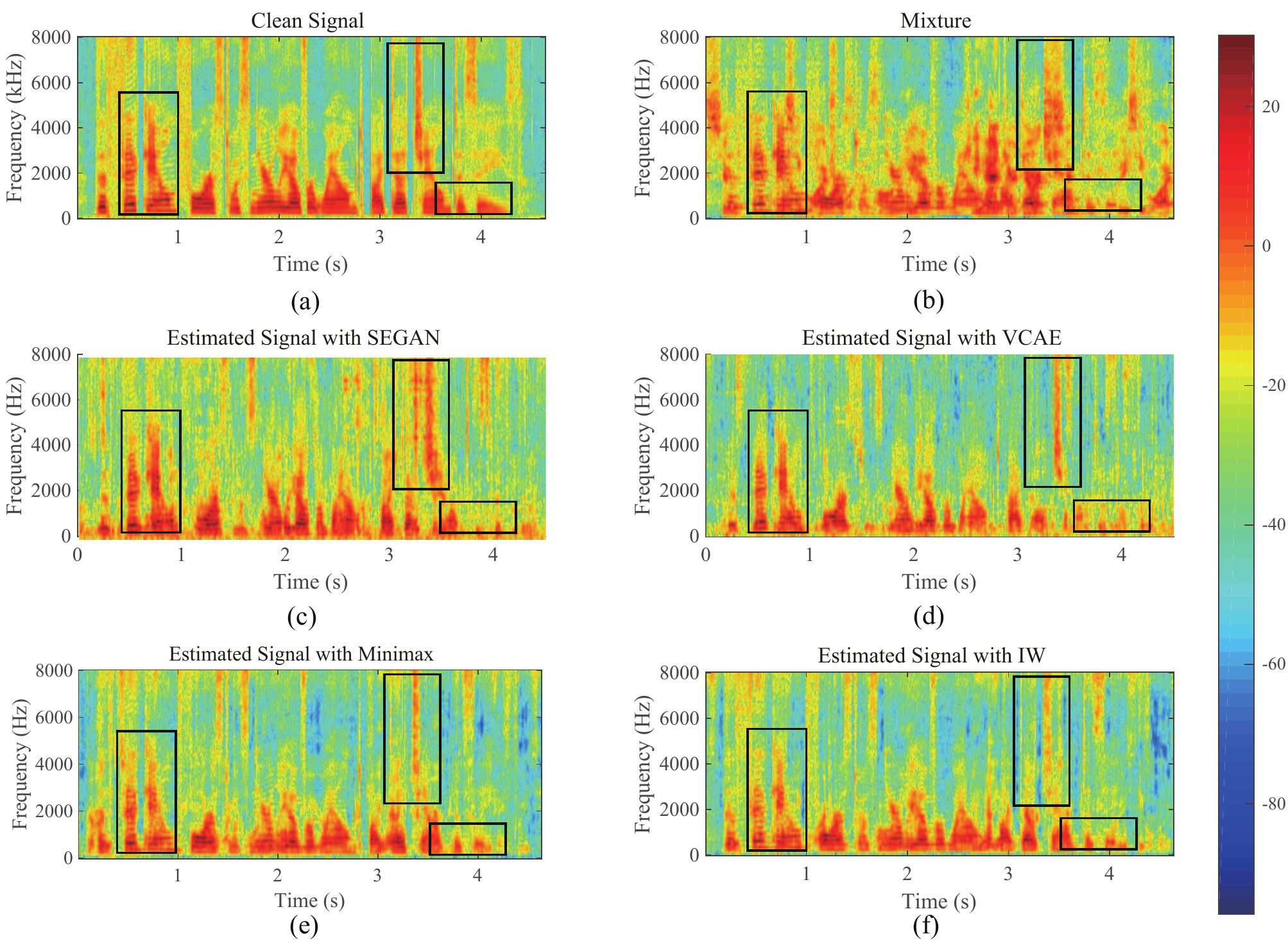}
\caption{The spectra of different signals: (a) clean speech signal; (b) mixture; (c) estimated signal with SEGAN ; (d) estimated signal with VCAE; (e) estimated signal with proposed Minimax (f) estimated signal with proposed IW. The mixture is generated with $station$ and -5 dB SNR level. The colormap indicates the relative power density.}\centering
\end{figure}
It can be observed from Fig. 2 that the noise interferences are better removed by the proposed methods from Fig. 2(e) and (f) compared to the baselines. Although the SEGAN and the original VCAE methods are competitive in speech enhancement, these both rely on the similarity between the source and target domains, and have limitations in transferring a well-trained model from one task or setting to another. Besides, SEGAN utilizes the complex neural network architecture to address the supervised speech enhancement. Therefore, the proposed methods take advantage of solving unsupervised speech enhancement and domain adaptation problems.

The above experimental results confirm that the proposed methods can further improve the speech enhancement and domain adaptation performance compared to the state-of-the-art methods, moreover, the improvement is statistically significant. The reason is that the proposed IW-VCAE method utilizes two classifiers to obtain the importance weights of the source samples and reduce the Jensen-Shannon divergence, respectively. Furthermore, the proposed minimax method maximizes the variables and train the classifier to minimize the risk under the worst-case. Therefore, when the weights of the target domain features are unknown, the desired speech signals are estimated more accurately by the proposed methods.
\section{CONCLUSIONS AND FUTURE WORK}
In this paper, the domain adaptation method was exploited to address the unsupervised speech enhancement problem. An IW scheme based on the classifiers in the networks was proposed to classify the source domain samples from the outlier weights and reduce the shift between the source and target domains. In Section III-C-2, speech enhancement performance of the proposed IW method had 3.9 $\%$, 11.0 $\%$, and 50.2 $\%$ improvements as compared to the original VCAE method in terms of three performance measurements. Moreover, the minimax method was proposed for the worse-case weights. Similarly, speech enhancement performance of the proposed minimax method had 4.4 $\%$, 12.6 $\%$, and 32.7 $\%$ improvements as compared to the original VCAE method. Thus, the experimental results confirmed that the speech enhancement and domain adaptation performances were improved by the proposed methods than the state-of-the-art approaches with the IEEE and TIMIT datasets. At more challenging scenarios in which the target domains were richer than the source domains, the minimax method would be the first choice.

For future work, the first direction is to explore the state-of-the-art neural networks such as the fully-convolutional time-domain audio separation network (Conv-TasNet) \cite{conv} and attention for the further improvement \cite{attention}. The speech enhancement and domain adaptation performance will be evaluated and compared to different networks. The second direction is exploiting new transfer learning algorithms to evaluate and further improve the domain adaptation performance.

\bibliographystyle{IEEEtran}
\bibliography{trans}
\begin{IEEEbiography}
    [{\includegraphics[width=1in,height=1.75in,clip,keepaspectratio]{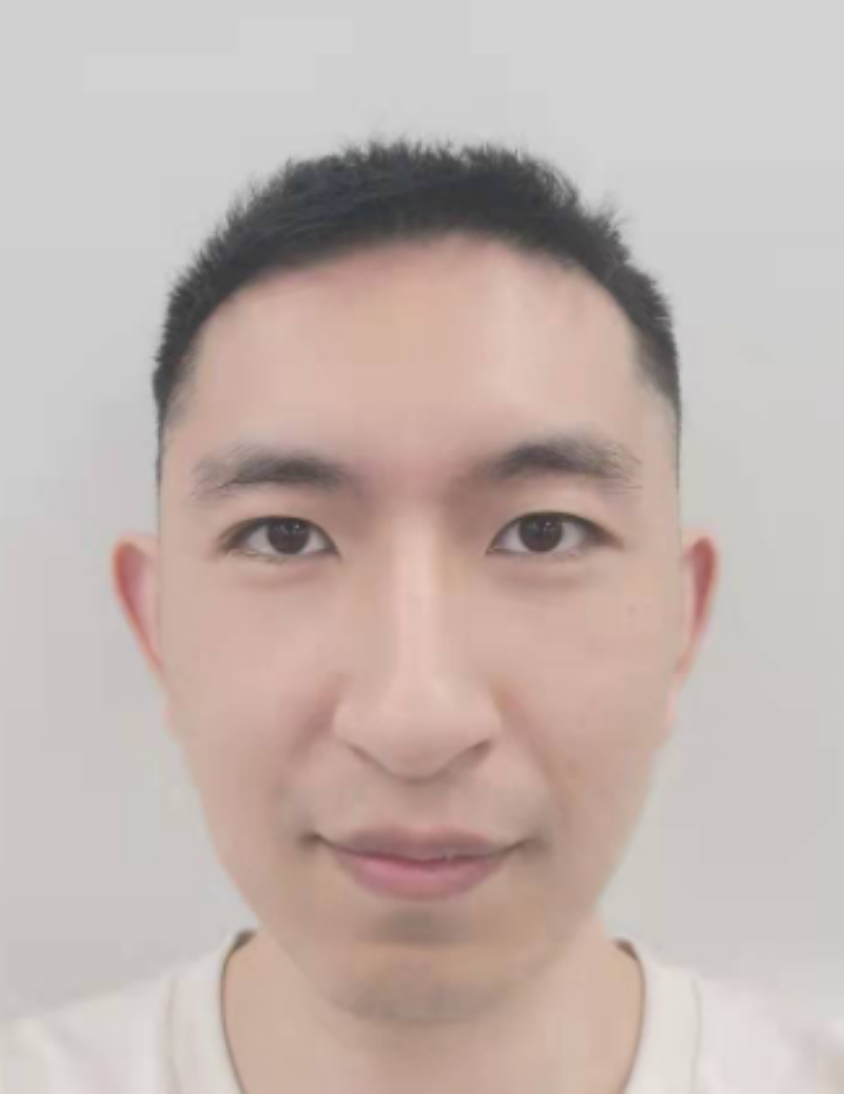}}]{Yi Li (Student Member, IEEE)}
received the B.Sc. degree in 2017, from University of Electronic Science and Technology of China and M.Sc degree in 2018 from Newcastle University, U.K, respectively. He is currently pursuing the Ph.D. degree within Intelligent Sensing and Communications Research Group, School of Engineering, Newcastle University, U.K. His research areas of interest include audio signal processing, speech source separation and enhancement based on deep learning.

\end{IEEEbiography}
\vskip -1pt plus -1fil
\begin{IEEEbiography}
    [{\includegraphics[width=1in,height=1.25in,clip,keepaspectratio]{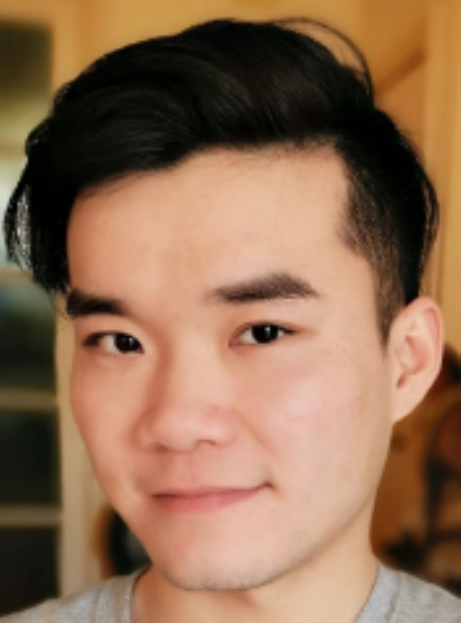}}]{Yang Sun (Member, IEEE)} received his Ph.D. degree within Intelligent Sensing and Communications (ISC) Research Group, School of Engineering, Newcastle University, U.K in 2019. Currently, Yang is a postdoctoral researcher at the Big Data Institute, University of Oxford. His research areas of interest include audio signal processing and medical image processing based on deep learning.
\end{IEEEbiography}
\vskip -2pt plus -1fil
\begin{IEEEbiography}
    [{\includegraphics[width=1in,height=1.75in,clip,keepaspectratio]{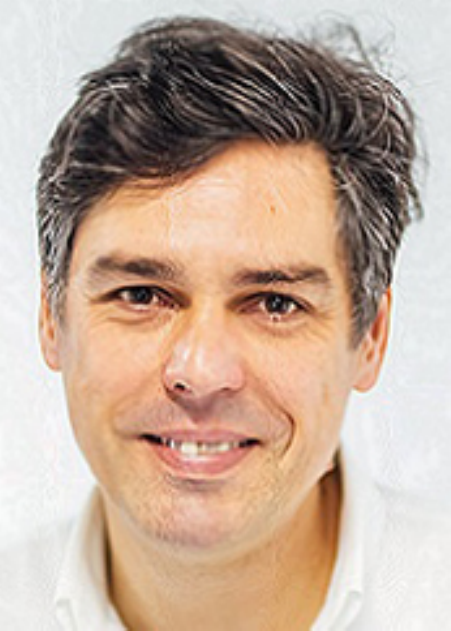}}]{Kirill Horoshenkov (Senoir Member, IEEE)}
is a Professor of Acoustics in the Department of Mechanical Engineering at the University of Sheffield, UK. His expertise is in sound propagation and acoustic sensing. He leads the UK Acoustics Network (www.acoustics.ac.uk). He is a Fellow of the Royal Academy of Engineering, UK Institute of Acoustics and Acoustical Society of America (ASA). He is author/co-author of over 100 papers in refereed journals and 10 patents. He was awarded the Tyndall Medal by the Institute of Acoustics in 2006 for his contribution to acoustics. One of his theoretical models for sound propagation in porous media is incorporated in Comsol and Altair commercial packages.
\end{IEEEbiography}
\vskip 0pt plus -1fil
\begin{IEEEbiography}
[{\includegraphics[width=1in,height=1.25in,clip,keepaspectratio]{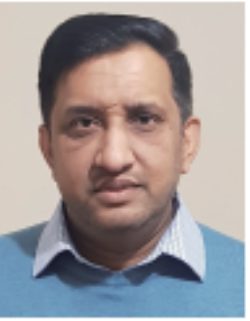}}]{Syed Mohsen Naqvi (Senoir Member, IEEE)}
received the Ph.D. degree in Signal Processing from Loughborough University, Loughborough, U.K., in 2009 and his Ph.D. thesis was on the EPSRC U.K. funded project. He was a Postdoctoral Research Associate on the EPSRC U.K.-funded projects and Research Excellence Framework (REF) Lecturer from 2009 to 2015. Prior to his postgraduate studies in Cardiff and Loughborough Universities U.K., he served the National Engineering and Scientific Commission (NESCOM) of Pakistan from 2002 to 2005.

Dr Naqvi is Associate Professor/Senior Lecturer in Signal and Information Processing at the School of Engineering, Newcastle University, Newcastle, U.K. He is leading Intelligent Sensing Lab at Newcastle University U.K. with major research focused on multimodal processing for human behavior analysis, multi-target tracking, mental health detection and speech processing; all for AI. He organized special sessions in FUSION, delivered seminars and was a speaker at UDRC Summer Schools 2015-2017. He has 150 publications with the main focus of his research being on audio-visual signal and information processing, machine learning and perception, reliable artificial intelligence, and action recognition and anomaly detection. He is an Associate Editor for Elsevier Journal on Signal Processing. He is Fellow of the Higher Education Academy. He is an Associate Editor for IEEE Transactions on Signal Processing. He is an Associate Editor for IEEE/ACM Transactions on Audio, Speech, and Language Processing.

\end{IEEEbiography}
\end{document}